\documentclass[3p,times,procedia]{elsarticle}
\usepackage{nupha_ecrc}

\volume{00}

\firstpage{1}

\journalname{Nuclear Physics A}

\runauth{}

\jid{nupha}

\jnltitlelogo{Nuclear Physics A}

\usepackage{amssymb}

\usepackage[figuresright]{rotating}

\graphicspath{
	{./}
	{./figures/}
}

\usepackage{amsmath}
\usepackage{amssymb}
\usepackage{graphicx}

\newcommand{\ii}{\text{i}}
\newcommand{\tave}[1]{\langle\!\langle{#1}\rangle\!\rangle}
\DeclareMathOperator{\im}{Im}

\begin{document}

\begin{frontmatter}



\dochead{XXVIIIth International Conference on Ultrarelativistic Nucleus-Nucleus Collisions\\ (Quark Matter 2019)}

\title{Electromagnetic and weak probes: theory}


\author{Ralf-Arno Tripolt}

\address{Institute for Theoretical Physics, Goethe University, Max-von-Laue-Str. 1, D-60438 Frankfurt am Main, Germany}

\begin{abstract}
An overview of the latest theoretical developments and results on electromagnetic and weak probes in relativistic heavy-ion collisions is presented. The possibilities to use electromagnetic probes, i.e., photons and dileptons, as a spectrometer, thermometer, chronometer, polarimeter, barometer, and multimeter of the collision system, as well as to explore the QCD phase diagram and to locate phase transitions such as the critical endpoint, are discussed.
\end{abstract}


\end{frontmatter}


\section{Introduction}
\label{sec:introduction}

Relativistic heavy-ion collisions allow to create an extreme state of matter that last existed in Nature a few microseconds after the Big Bang. They are furthermore the only means by which the bulk properties of a nonabelian gauge theory, such as Quantum Chromodynamics (QCD), can be compared to experiment. This is particularly important in a situation where the coupling strength is not small, implying a wealth of nonperturbative and collective phenomena which can only be unraveled in close collaboration between theory and experiment. Such experiments are currently performed at the SPS and the LHC at CERN, RHIC at BNL, SIS at GSI, as well as at future facilities such as the Facility for Antiproton and Ion Research (FAIR), the Nuclotron-based Ion Collider fAcility (NICA), the High Intensity heavy ion Accelerator Facility (HIAF), and the heavy-ion program at the Japan Proton Accelerator Complex (J-PARC).

Electromagnetic and weak probes enjoy a unique status in heavy-ion collisions since they do not interact strongly with the created medium and can therefore carry information directly to the detectors. Weak probes, i.e., $W^{\pm}$ and $Z^0$ gauge bosons, are produced by initial hard scattering processes and can be used to learn about initial-state effects such as nuclear modifications of parton distribution functions (PDFs). Electromagnetic probes, i.e., photons and dileptons, are produced at all stages of the collision and can be used to learn about the properties of the created hot and dense medium due to their soft and penetrating nature.

In the following, I review recent theoretical developments in the context of electromagnetic and weak probes with a particular emphasis on photons and dileptons and their application to learn about the temperature of the created medium, its lifetime, the degree of collectivity, in-medium spectral functions, chiral symmetry, changes in the degrees of freedom, transport coefficients, and phase transitions.

\section{Electromagnetic probes}
\label{sec:EM_probes}

The emission rates of photons and dileptons can be obtained either by using thermal field theory or relativistic kinetic theory.
In thermal field theory, the central role is played by the electromagnetic (EM) current-current correlation function, defined as
\begin{equation}
\Pi_{\rm EM}^{\mu \nu}(M,p;\mu_B,T)  = -\ii \int d^4x \ e^{ip\cdot x} \
\Theta(x_0) \ \tave{[j_{\rm EM}^\mu(x), j_{\rm EM}^\nu(0)]},
\end{equation}
where $\langle \langle \cdots \rangle \rangle$ denotes the expectation value at finite temperature. The thermal emission rates of photons and dileptons are then given, respectively, by 
\begin{align} 
p_0 \frac{dR_\gamma}{d^3p}& = 
-\frac{\alpha_{\rm EM}}{\pi^2} \
f^B(p_0;T) \  \ g_{\mu\nu} 
\ \im \Pi_{\rm EM}^{\mu\nu} (M=0,p;\mu_B,T),
\label{eq:photon1} \\
\frac{dR_{ll}}{d^4p}& =
-\frac{\alpha_{\rm EM}^2}{\pi^3 M^2} \
f^B(p_0;T) \  \frac{1}{3} \ g_{\mu\nu} 
\ \im \Pi_{\rm EM}^{\mu\nu} (M,p;\mu_B,T),
\end{align}
where $f^B(p_0;T)=1/(e^{p_0/T}-1)$ is the Bose distribution function, see e.g.~\cite{McLerranToimela1985}. We note that photon and dilepton rates are governed by the same underlying object, i.e., the EM spectral function, $\im \Pi_{\rm EM} (M,p;\mu_B,T)$, albeit in different kinematic regimes. We also note that the leading order in the photon rate is $\mathcal{O}(\alpha_s)$ while for the dilepton rate we have $\mathcal{O}(\alpha_s^0)$. 

Alternatively, the emission rate of (virtual) photons can be expressed using relativistic kinetic theory,
\begin{align}
p_0 \frac{dR}{d^3p} =
\int \frac{d^3q_1}{2(2\pi)^3E_1}
\frac{d^3q_2}{2(2\pi)^3E_2}\frac{d^3q_3}{2(2\pi)^3E_3}
(2\pi)^4
\delta^{(4)}(q_1+q_2\rightarrow q_3+p)
\left|{\cal M}\right|^2\frac{f(E_1)f(E_2)[1\pm f(E_3)]}{2(2\pi)^3},
\end{align}
where $f(E_i)$ are the distribution functions of the associated particles and ${\cal M}$ is the invariant scattering matrix element, see e.g.~\cite{KajantieKapustaMcLerranEtAl1986}. This microscopic formulation is well suited for non-equilibrium calculations and processes at high momenta, e.g., perturbation theory, while medium effects are more readily implemented within the thermal field-theory approach.

\subsection{Photons}
\label{sec:photons}

Photons are produced at all stages of a heavy-ion collision and can be classified into decay photons and direct photons. Decay photons originate from the decay of long-lived resonances such as pions, eta, and omega mesons after freeze-out, while all other photons are called direct photons. Experimentally, the direct-photon contribution is obtained by subtraction of the decay-photon contribution from the inclusive (total) spectra. Direct photons can be classified as follows:
\begin{itemize}
	\item hard photons (with high transverse momenta $p_T$) from initial hard-scattering processes (`prompt'), jet-fragmentation processes, and vacuum bremsstrahlung
	\item pre-equilibrium photons
	\item photons from jet-medium interaction, e.g.~jet-$\gamma$ conversion and medium-induced bremsstrahlung
	\item thermal photons from the QGP and the hot and dense hadron-gas phase (including short-lived resonances like $\omega$, $a_1$, $\Delta$, $N^*$, ...)
	\item other sources: hadronic bremsstrahlung, $B$-field induced photons, ...
\end{itemize}
Thermal photons from the QGP and the hadron gas are of particular interest, since they contain information on the hot and dense equilibrium phase of the collision. At high temperatures and momenta, the photon-emission rate is well-described by perturbative QCD (pQCD). The corresponding result was presented in \cite{ArnoldMooreYaffe2001} at leading order (LO), $\mathcal{O}(\log(g))$, and in \cite{GhiglieriHongKurkelaEtAl2013} at next-to-leading order (NLO), $\mathcal{O}(g)$.
Both results are still widely used to calculate the thermal-photon contribution from the QGP phase.

Another possibility to calculate thermal-photon rates from first principles is given by lattice QCD (lQCD). The main challenge here is to extract the EM vector spectral function from numerical data on the corresponding propagator, 
\begin{align}
G^E_V(\tau,\vec{p})=\int_0^{\infty}\frac{dp_0}{2\pi} \im \Pi_{\rm EM} (p_0,\vec{p})
\frac{\cosh[p_0(\tau-1/2T)]}{\sinh(p_0/2T)}.
\end{align}
Recent lQCD results for the thermal QGP photon rate were presented in \cite{GhiglieriKaczmarekLaineEtAl2016} and \cite{BrandtFrancisHarrisEtAl2018}. In \cite{GhiglieriKaczmarekLaineEtAl2016}, lattice results for the vector-current correlator for quenched QCD were analyzed with the help of a polynomial interpolation for the spectral function, which vanishes at zero frequency and matches to high-precision perturbative results at large invariant masses. In \cite{BrandtFrancisHarrisEtAl2018}, the Euclidean vector-current correlation function was computed for dynamical QCD with two flavors and analyzed using the Backus-Gilbert method as well as a model ansatz. The obtained results are in agreement with expectations from perturbation theory, however, the uncertainties remain large. Future lQCD calculations are therefore necessary, for example using larger lattices or higher statistics, in order to obtain photon rates with small uncertainties that can be used in comparisons to experimental data.

The thermal radiation from the hot and dense hadron gas phase is usually obtained from effective model calculations, using both thermal field-theory techniques as well as relativistic kinetic theory. As for example discussed in \cite{PaquetShenDenicolEtAl2016}, different hadronic contributions need to be taken into account in order to arrive at a complete picture. Photons originating from thermal reactions of mesonic origin were for example calculated in \cite{TurbideRappGale2004} using a Massive Yang-Mills (MYM) approach, where vector and axial-vector mesons are introduced as massive gauge fields of, e.g., a chiral $U(3)$ symmetry. One particular advantage of the MYM approach is that it is capable of yielding adequate hadronic phenomenology at tree level with a rather limited set of adjustable parameters. Using this approach, the photon-emission rate of a hot meson gas consisting of $\pi$, $K$, $K^*$, $\rho$, and $a_1$ mesons was obtained \cite{TurbideRappGale2004}. Therein, also the baryonic photon contribution was calculated, using hadronic many-body theory. This framework is based on effective hadronic Lagrangians with constant parameters that are constrained by empirical information. In this way, the in-medium $\rho$-meson propagator can be calculated, which then determines the EM spectral function by using vector-meson dominance,
\begin{align}
\text{Im}\Pi_{\text{EM}}(M,p;\mu_B,T)\approx\frac{m_\rho^4}{g^2}\text{Im}D_\rho(M,p;\mu_B,T).
\end{align}
When evaluated at zero invariant mass, this procedure accounts for the baryonic photon contributions, be it radiative decays or reactions of the type $\pi N \rightarrow \pi N \gamma$, $N N \rightarrow N N\gamma$, where $N$ represents a nucleon. In addition, also the contributions from $\pi\pi$ bremsstrahlung, as obtained in \cite{HeffernanHohlerRapp2015}, and from the reactions $\pi\rho\rightarrow \omega \gamma$, $\pi\omega\rightarrow \rho \gamma$, and $\pi\omega\rightarrow \rho \pi$, as presented in \cite{HoltHohlerRapp2016}, need to be taken into account. When comparing the total hadronic photon-production rate with the QGP rate near the crossover temperature, one finds that they are very similar, giving rise to the so-called duality of partonic and hadronic rates.

\begin{figure}[t!]
	{\includegraphics[width=0.49\textwidth]{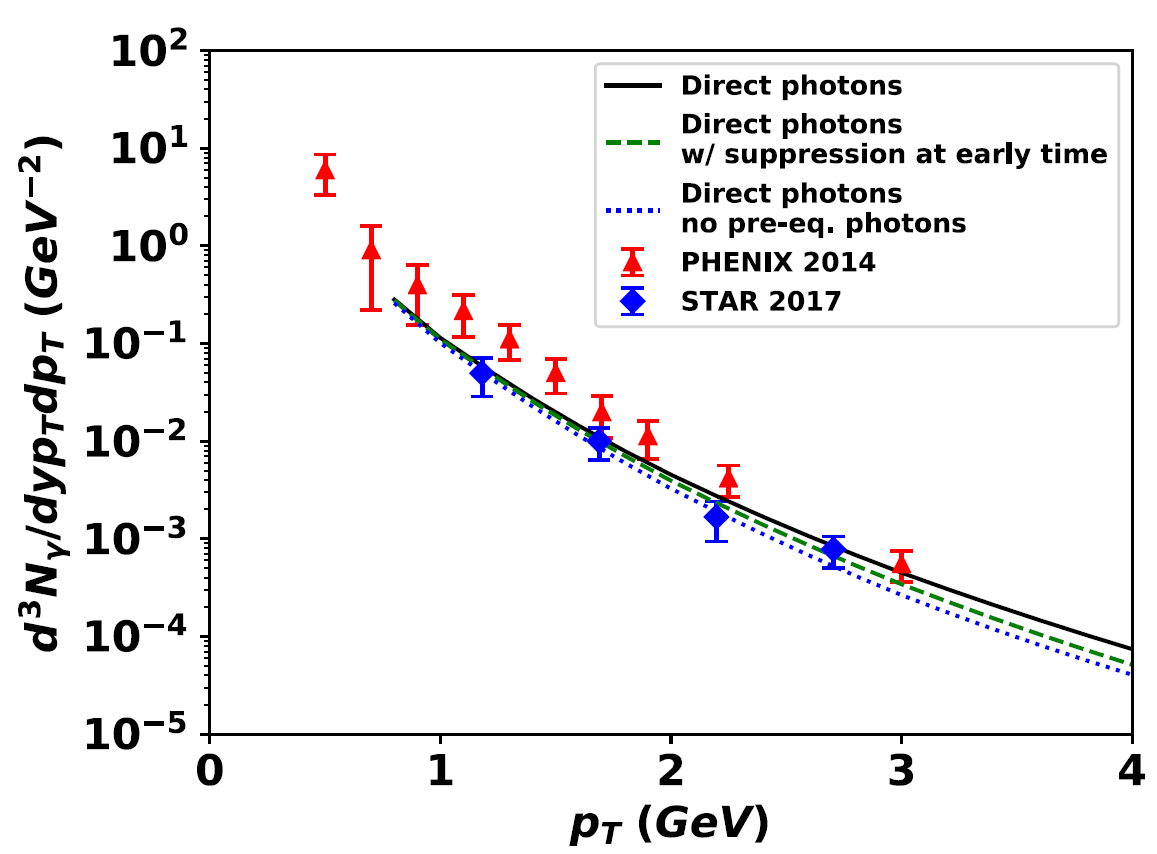}}\hspace{0.02\textwidth}
	{\includegraphics[width=0.49\textwidth]{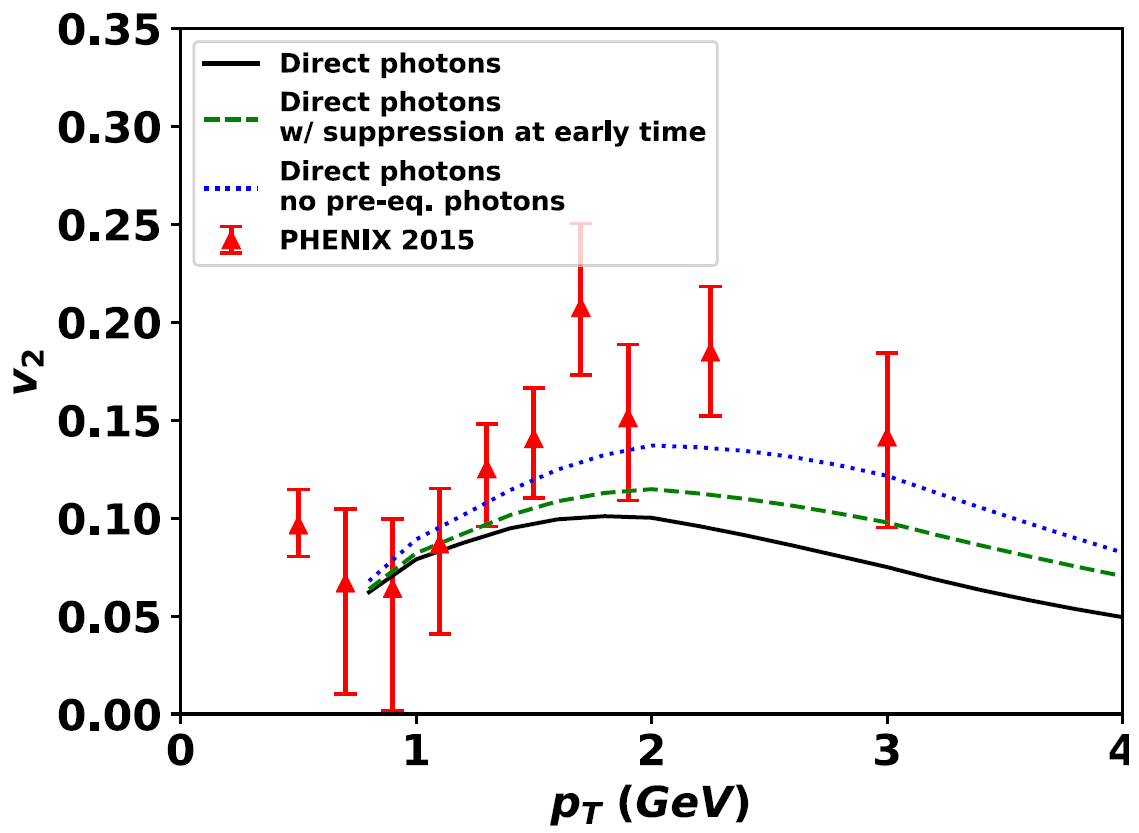}}
	\caption{Comparison of experimental data on the direct-photon spectrum (left) and the elliptic-flow coefficient $v_2$ (right) as measured at RHIC by the PHENIX collaboration \cite{Adareothers2016} and the STAR collaboration \cite{Adamczykothers2017} at $\sqrt{s_{\text{NN}}}=200$~GeV for centralities of $20-40\%$ to recent theoretical results, see text for details. Plots taken from \cite{Gale2019}.}
	\label{fig:v2}
\end{figure}

Together with the contributions from other sources, such as prompt photons and non-cocktail hadronic decay photons, the photon rates need to be integrated over the space-time evolution of the collision process in order to obtain the measured photon spectra. Recent results for direct-photon spectra at RHIC and LHC energies have been obtained in \cite{PaquetShenDenicolEtAl2016}, where a state-of-the-art event-by-event hydrodynamical model with IP-Glasma initial states was used. An update of these calculations was presented at this conference by C.~Gale, see Fig.~\ref{fig:v2}. In this calculation, the dynamical evolution was supplemented with a K\o MP\o ST phase, i.e., an effective kinetic-theory approach to the pre-hydro phase \cite{KaymakcalanRajeevSchechter1984}. While the experimental results for the direct-photon spectrum as measured at RHIC by the PHENIX collaboration \cite{Adareothers2016} and the STAR collaboration \cite{Adamczykothers2017} at $\sqrt{s_{\text{NN}}}=200$~GeV for centralities of $20-40\%$ show some deviation from each other, the theoretical results agree very well with the STAR data. However, within the uncertainties, all results are compatible, also for the elliptic-flow coefficient $v_2$, as shown in the right panel of Fig.~\ref{fig:v2}. We note that good agreement between theory and data is also be obtained at LHC energies, see for example \cite{PaquetShenDenicolEtAl2016}.

At photon energies below 2-3 GeV, the measured photon spectra are approximately exponential and can be characterized by their inverse logarithmic slope, often called ``effective temperature''. This effective temperature is, however, blue-shifted due to the transverse flow of the medium, see \cite{HeesGaleRapp2011} where the quantitative blueshift effect and the issue of the "true" temperature was first pointed out. By modelling the evolution of the radiating medium hydrodynamically, the relation between the effective temperature and the true temperature of the fireball has recently been investigated in \cite{ShenHeinzPaquetEtAl2014}. It was found that at RHIC and LHC energies most photons are emitted from fireball regions with temperatures near the quark-hadron phase transition, but that their effective temperature is significantly enhanced by strong radial flow, see also \cite{HeesGaleRapp2011}. This finding, i.e.~that a large part of the photons comes from near $T_c$ and the hadronic phase, is essential for solving the so-called ``$v_2$-puzzle".

Photons are also useful as a ``viscometer'', see e.g.~\cite{ShenHeinzPaquetEtAl2015}, where viscous photon emission from nuclear collisions at RHIC and LHC was investigated by evolving fluctuating initial density profiles with event-by-event viscous hydrodynamics. Momentum spectra of thermal photons, radiated by these explosively expanding fireballs, and their $p_T$-differential anisotropic flow coefficients were computed, both with and without accounting for viscous corrections to the standard thermal emission rates. The overall effect of viscous corrections to the rates on the direct photon spectra was found to be small, which can be understood from the fact that viscous corrections are larger at higher $p_T$, where prompt photons dominate over thermal ones. The direct photon $v_2$, on the other hand, is suppressed at higher $p_T$ by both shear and bulk corrections to the photon rates, with the suppression being of the order of $20-30\%$ \cite{PaquetShenDenicolEtAl2016}.

\subsection{Dileptons}
\label{sec:dileptons}

Similar to photons, dileptons, i.e., $e^+e^-$ or $\mu^+\mu^-$ pairs, are produced at all stages of the collision and can be classified as follows:
\begin{itemize}
	\item `primordial' dileptons from $q\bar{q}$ annihilation, i.e., Drell-Yan processes
	\item thermal dileptons from the QGP and the hot and dense hadron-gas phase (including dileptons from multi-meson reactions and decays of short-lived resonances like $\rho$, $\omega$, $a_1$, $\Delta$, $N^*$, ...)
	\item dileptons from decays of long-lived mesons and baryons (e.g. from $\pi^0$, $\eta$, $\phi$, $J/\Psi$, $\Psi$', $D\bar D$, ...)
\end{itemize}
In the following we will again focus on the thermal radiation from the QGP and the hot and dense hadron-gas phase, which is most readily described by the thermal field-theory approach utilizing the EM spectral function. In the vacuum, the EM spectral function is well-known from the inverse process of $e^+e^-$ annihilating into hadrons, $R=\sigma(e^+ e^- \to {\rm hadrons})/\sigma(e^+ e^- \to\mu^+ \mu^-) \propto  \im\Pi_{\rm EM}^{\rm vac}/M^2$. In the low-mass regime (LMR), $M\leq 1$~GeV, one finds that the EM spectral function is saturated by the spectral functions of the light vector mesons, cf.~the aforementioned vector-meson dominance, while at higher energies it is well described by quark degrees of freedom. 
The main challenge for a realistic description of thermal-dilepton rates at finite temperature and density is therefore the calculation of the in-medium vector-meson spectral functions. This can be achieved using different frameworks. One possibility is given by hadronic many-body theory, which is based on effective hadronic Lagrangians and allows to calculate the in-medium modifications of the corresponding propagator, see, e.g., \cite{Rapp:1999ej,Rapp2011}. In particular the Rapp-Wambach spectral functions have proven to be very successful in the description of experimental data and are still widely used. 

\begin{figure}[t!]
	\includegraphics[width=0.49\textwidth]{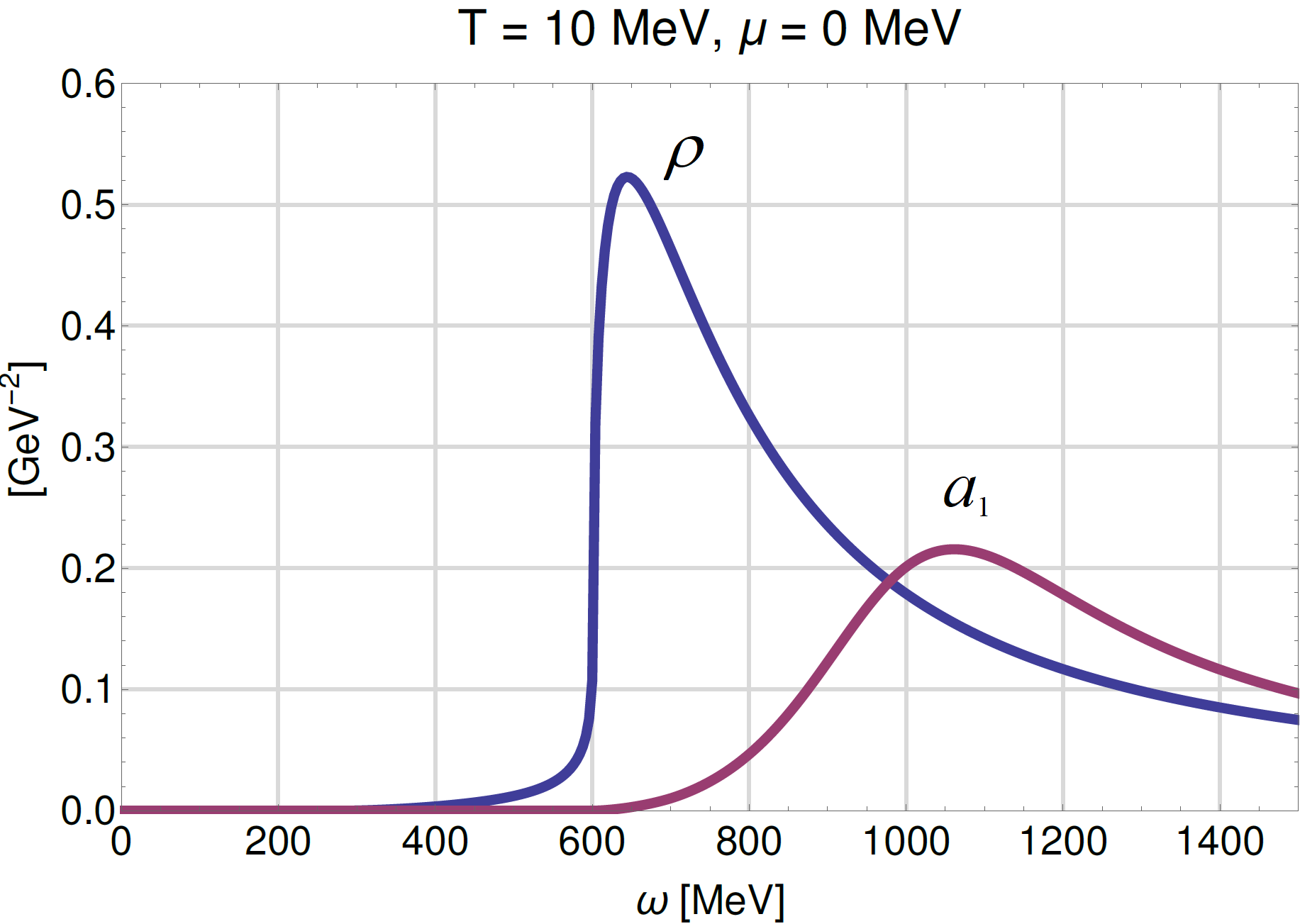}\hspace{0.02\textwidth}
	\includegraphics[width=0.49\textwidth]{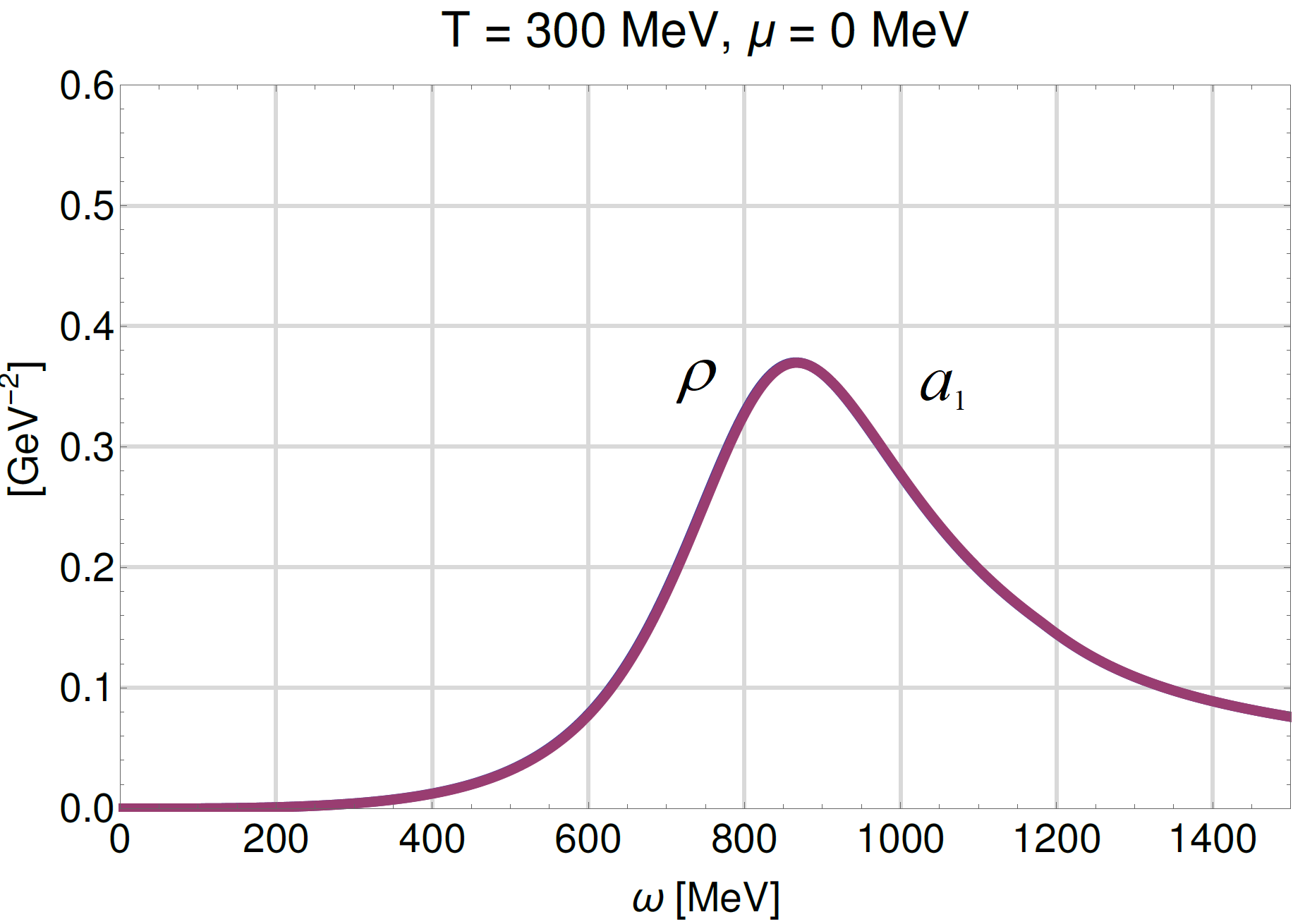}
	\caption{The $\rho$ and $a_1$ spectral functions as obtained from the FRG calculation of \cite{JungRenneckeTripoltEtAl2017} are shown at a temperature of $T=10$~MeV (left) and at $T=300$~MeV (right) for vanishing chemical potential. While at $T=10$~MeV chiral symmetry is spontaneously broken and the spectral functions are close to their vacuum values, at $T=300$~MeV the spectral functions become completely degenerate, as expected due to the restoration of chiral symmetry.}
	\label{fig:FRG}
\end{figure}

Recently, the in-medium $\rho$ spectral function was also calculated using the Functional Renormalization Group (FRG) approach \cite{JungRenneckeTripoltEtAl2017,JungSmekal2019} which is a powerful non-perturbative framework that implements Wilson's coarse-graining idea by integrating out quantum fluctuations from high to low scales successively. In particular, the FRG is able to compute thermodynamical and spectral properties on the same footing and also properly deals with phase transitions at finite temperature and density. Recent results also show that FRG results are now in quantitative agreement with results from lQCD, for example when comparing the chiral quark condensate \cite{FuPawlowskiRennecke2019}. In \cite{JungRenneckeTripoltEtAl2017}, an effective low-energy theory for QCD motivated by the gauged linear sigma model was used to obtain the $\rho$ and the $a_1$ spectral function at finite temperature and density, see also Fig.~\ref{fig:FRG}. It was shown that these spectral functions become degenerate near the chiral crossover temperature and that the $a_1$ mass moves towards the $\rho$ mass, while the latter stays approximately constant. These spectral functions have also been used to obtain a first estimate of the expected dilepton rates, in particular near the critical endpoint in the corresponding phase diagram \cite{TripoltJungTanjiEtAl2019}. However, additional improvements such as the inclusion of baryonic degrees will be necessary before a quantitative comparison to experimental data can be made.

A similar result on in-medium $\rho$ and $a_1$ spectral functions was found in \cite{Hohler:2013eba}, where a combined analysis of finite-temperature QCD and Weinberg sum rules has been carried out. Together with calculated in-medium $\rho$ spectral functions, viable in-medium $a_1$ spectral functions were searched for that satisfy both QCD and Weinberg sum rules within their typical accuracy of $\sim 0.5\%$. A solution was found that gradually becomes degenerate with the vector-meson spectral function, again suggesting a mechanism of chiral restoration by which the broadening of both $\rho$ and $a_1$ is accompanied by a reduction of the $a_1$ mass moving toward the $\rho$ mass.

A microscopic investigation of the in-medium $\pi-\rho-a_1$ system has recently been conducted within the Massive-Yang Mills (MYM) framework \cite{Hohler:2015iba}. Therein, the difficulties of the MYM approach to describe the vacuum $a_1$ spectral function could be overcome by introducing a broad $\rho$ propagator into the $a_1$ self-energy, accompanied by vertex corrections. The results obtained at finite temperature (pion gas) are similar to the phenomenological sum-rule analysis and corroborate the ``burning off" of the chiral mass splitting as a mechanism of chiral degeneracy.

Realistic in-medium spectral functions are key for the description of dilepton rates in heavy-ion collisions and a clear interpretation of the results. In \cite{RappHees2016}, it was shown how the predictions of hadronic many-body theory for a melting $\rho$ meson, coupled with QGP emission utilizing a modern lQCD-based equation of state, yield a quantitative description of dilepton spectra in heavy-ion collisions at the SPS and the RHIC beam-energy scan program. It was moreover shown that the integrated low-mass excess radiation is related to the fireball lifetime, i.e., represents a chronometer, and that the slope of the invariant-mass spectrum at intermediate energies can be used to extract an effective temperature, i.e., represents a thermometer. 

\begin{figure}[t!]
	\includegraphics[width=0.49\textwidth]{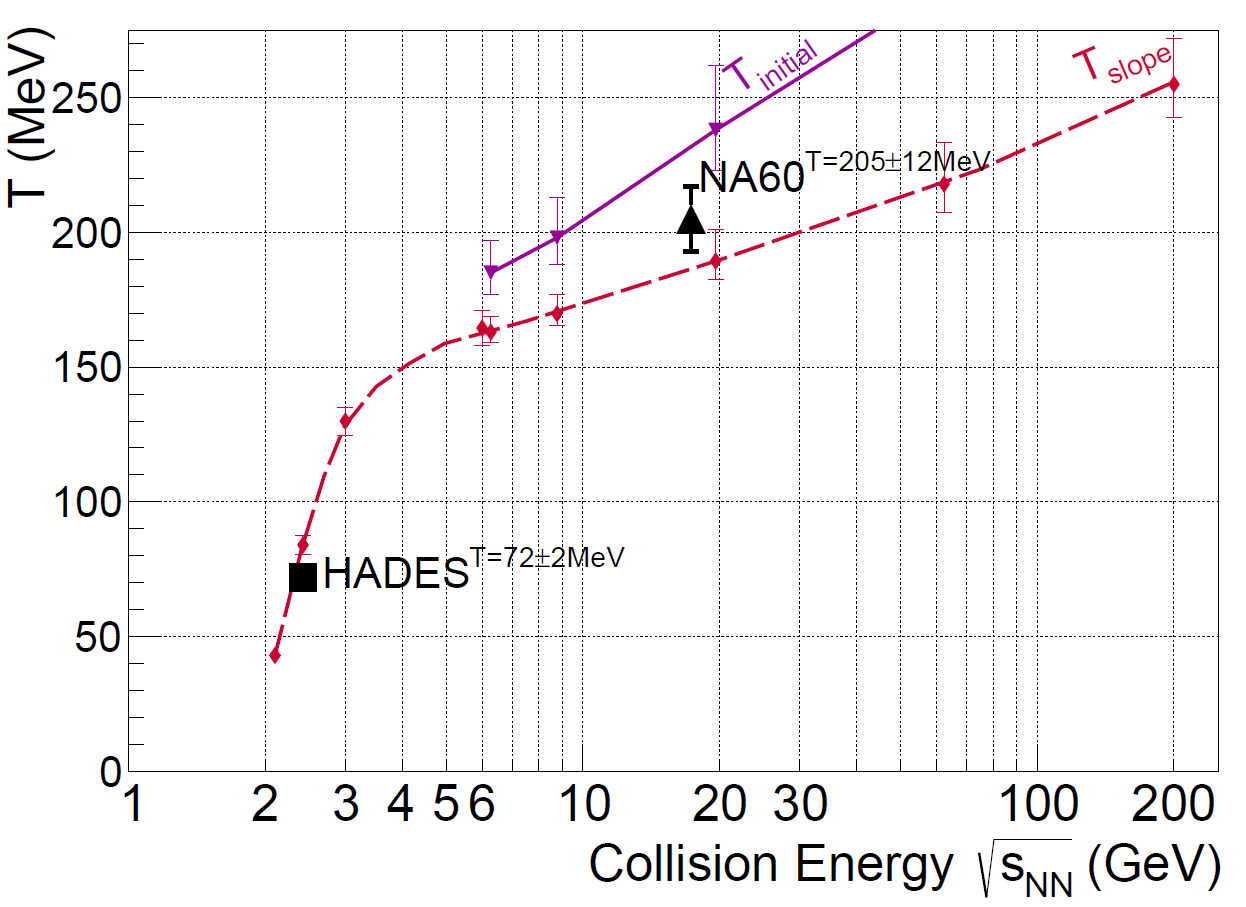}\hspace{0.02\textwidth}
	\includegraphics[width=0.49\textwidth]{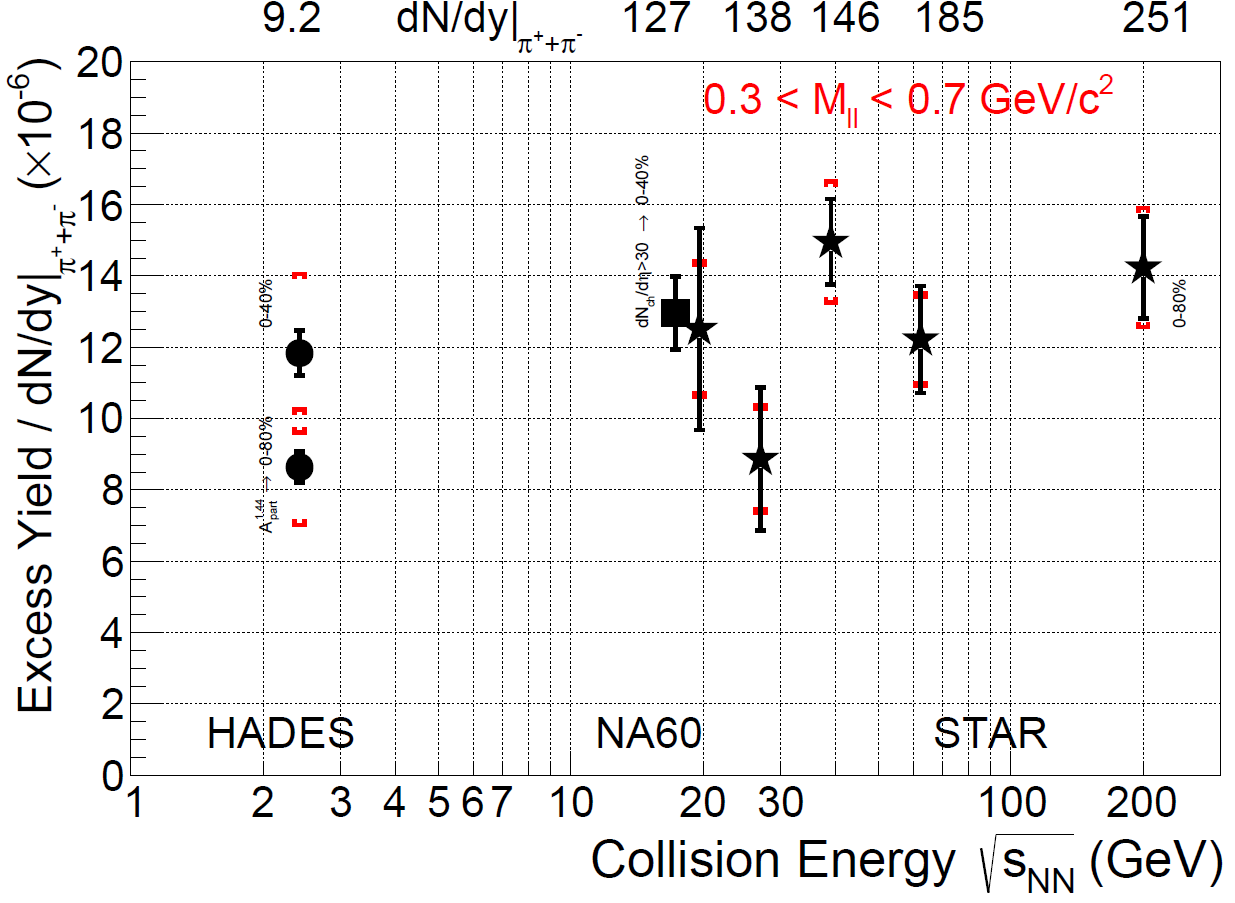}
	\caption{Left: Excitation function of the inverse slope parameter, $T_\text{\text{slope}}$, and of the initial temperature, $T_\text{initial}$, as obtained from a fireball model \cite{RappHees2016} and from a coarse-grained transport approach \cite{GalatyukHohlerRappEtAl2016} in comparison to temperatures extracted from dilepton spectra measured by HADES \cite{Adamczewski-Muschothers2019} and NA60 \cite{Specht2010}. Right: Excitation function of the integrated low-mass dilepton excess yields as measured by HADES, NA60, and STAR \cite{Galatyuk}. Plots compiled by T.~Galatyuk.}
	\label{fig:CEP}
\end{figure}

In the left panel of Fig.~\ref{fig:CEP}, the excitation function of the inverse slope parameter, $T_\text{\text{slope}}$, and of the initial temperature, $T_\text{initial}$, as obtained from a fireball model \cite{RappHees2016} and from a coarse-grained transport approach \cite{GalatyukHohlerRappEtAl2016} obtained in \cite{RappHees2016}, are compared to temperatures extracted from dilepton spectra measured by HADES \cite{Adamczewski-Muschothers2019} and NA60 \cite{Specht2010}. Since the model calculation used to obtain $T_\text{\text{slope}}$ does not incorporate any phase transition, any deviation from the shown behavior may indicate a phase transition. Indeed, a first-order transition is expected to show up as a plateau in the excitation function of $T_\text{\text{slope}}$, similar to a caloric curve, where latent heat has to be ``burned off''. In the right panel of Fig.~\ref{fig:CEP}, the integrated low-mass excess yield is shown as obtained from HADES, NA60, and STAR \cite{Galatyuk}. These data points reflect the life-time of the system, which gradually increases towards higher collision energies. Also this observable can be used to indicate a phase transition. In particular, a critical point is expected to show up as a peak in this excitation function since extra radiation will be generated when the system lives longer around such a second-order phase transition.

As shown in \cite{SperanzaJaiswalFriman2018}, dileptons can also be used as a polarimeter, since their angular distribution in the photon rest frame is related to the polarization of the initial virtual photon. In particular, anisotropy coefficients have been calculated for quark-antiquark annihilation in the QGP and pion annihilation in the hadronic phase for a static medium in global equilibrium and for a longitudinally expanding system. It was shown in \cite{SperanzaJaiswalFriman2018} that a polarization is to be expected even in thermal equilibrium and that this polarization can be used to learn about the underlying production mechanisms of dileptons. 

Another interesting application of dileptons is to use them to measure transport coefficients. While the influence of dissipative corrections, i.e., effects from shear and bulk viscosity, on dilepton production was for example studied in \cite{VujanovicDenicolLuzumEtAl2018, VujanovicPaquetShenEtAl2019}, there are currently no calculations of the vector spectral density including baryon interactions, off-shell effects, and viscous corrections. The electrical conductivity, on the other hand, can be directly obtained as the low-energy limit of the EM spectral function,
\begin{align}
\sigma_{\text{el}}=-e^2 \lim_{p_0\rightarrow 0} \frac{\partial}{\partial p_0} \text{Im} \Pi_{\text{EM}}(p_0,|\vec{p}|=0).
\end{align}
Recent results for the electrical conductivity have for example been presented in \cite{GreifGreinerDenicol2016, AtchisonRapp2017, GhoshMitraSarkar2018}. The large spread of results in the literature reflects the difficulty to obtain realistic results for a strongly interacting medium. However, the electrical conductivity can in principle also be measured experimentally by looking at the low-energy regime of dilepton spectra and the conductivity peak at small invariant masses.

\section{Weak probes}
\label{sec:weak_probes}

Weak probes, i.e., $W^\pm$ and $Z^0$ bosons, are produced by initial hard-scattering processes and can subsequently decay into leptons and neutrinos via the weak interaction. They can therefore be used to learn about cold nuclear matter effects and serve as a reference for medium-induced effects. In \cite{Acharyaothers2018}, for example, experimental results on $Z^0$ boson production at large rapidities in Pb-Pb collisions at $\sqrt{s_{\text{NN}}}=5.02$~TeV were compared to theoretical calculations. It was shown that the invariant yield and the nuclear modification factor $R_{AA}$ are well described by calculations that include nuclear modifications of the parton distribution functions $f_i^{p/A}(x,Q^2)$, as described by a nuclear modification factor $R_i^A(x,Q^2)$,
\begin{align}
f_i^{p/A}(x,Q^2)=R_i^A(x,Q^2)f_i^{p}(x,Q^2),
\end{align}
where $f_i^{p}(x,Q^2)$ represents the free proton PDF. While the predictions using vacuum PDFs deviate from the data, in particular the calculations obtained with the recent EPPS16 \cite{EskolaPaakkinenPaukkunenEtAl2017} parameterisation as well as with the nCTEQ15 \cite{KusinaLyonnetClarkEtAl2017} nuclear PDFs are in good agreement with the data.

\section{Summary}
\label{sec:summary}

Electromagnetic and weak probes enjoy a unique status in heavy-ion collisions since they do not interact strongly with the created medium and can provide undistorted information about their production site. In particular EM probes, i.e., photons and dileptons, are the only probes that are soft and penetrating and can therefore provide us with a wide range of insights on the properties of hot and dense QCD matter. Dileptons are especially useful since they have an additional `degree of freedom', i.e., the invariant mass. They can therefore provide basic kinematical information such as the fireball temperature, the degree of collectivity, and the lifetime, but also dynamical information on in-medium spectral functions encoding changes in degrees of freedom and chiral symmetry restoration as well as on transport coefficients like the electrical conductivity. 
In recent years, the melting of the $\rho$ meson in a strongly-interacting hadronic medium was confirmed by various experiments and theoretical calculations, indicating a transition from hadronic degrees of freedom towards a quark-antiquark continuum that is consistent with chiral restoration. There is also emerging consensus that chiral partners become degenerate at the ground state mass in a way that the chiral mass splitting burns off but the ground-state mass, which is then likely generated by another mechanism such as the gluon condensate, remains.
New theoretical developments, e.g.~from the Functional Renormalization Group or lattice QCD, are expected to provide chirally and thermodynamically consistent vector-meson spectral functions that will allow for a phenomenologically successful description of experimentally measured dilepton spectra while at the same time being well-founded in theory. Together with high-precision measurements expected from running and upcoming experiments such as STAR BES-II, NA60+, FAIR, NICA, HIAF, or J-PARC HI, this will allow to establish a clear connection to chiral symmetry restoration and eventually also to identify QCD phase transitions such as the critical endpoint.

\section*{Acknowledgements}
I would like to thank the organizers of the Quark Matter 2019 conference for the opportunity to give this overview talk. I would also like to thank Tetyana Galatyuk, Charles Gale, Frank Geurts, Hendrik van Hees, Ralf Rapp,  Dirk Rischke, Lorenz von Smekal, and Jochen Wambach for valuable discussions. The author is supported by  the  Deutsche Forschungsgemeinschaft (DFG, German Research Foundation)  project number 315477589  TRR 211, as well as by the BMBF under grant No.~05P18RFFCA.





\bibliographystyle{elsarticle-num}







\end{document}